\documentclass[aps,prl,showkeys,showpacs,superscriptaddress,reprint,amsmath,amssymb,twocolumn,floatfix,nofootinbib]{revtex4-1}
\usepackage[pdftex]{graphicx}
\usepackage{colortbl}
\usepackage{url}
\usepackage{lmodern}
\usepackage{hyperref}

\newcommand{\ind}[1]{\mathbf{1}_{#1}}
\newcommand{\bs}{S_\mathcal{B}}
\newcommand{\hbs}{\hat{S}_\mathcal{B}}
\newcommand{\br}[1]{\left( #1 \right)}
\newcommand{\eps}{\varepsilon}
\newcommand{\vF}{{F}}
\newcommand{\vx}{{x}}
\newcommand{\mR}{{R}} 
\newcommand{\mS}{{X}}
\newcommand{\mB}{{B}}
\newcommand{\mA}{{A}}

\begin{document}

\title{Potentials and Limits to Basin Stability Estimation} 

\author{Paul \surname{Schultz}}
\email{pschultz@pik-potsdam.de}
\affiliation{Potsdam Institute for Climate Impact Research, P.O. Box 60 12 03, 14412 Potsdam, Germany}
\affiliation{Department of Physics, Humboldt University of Berlin, Newtonstr. 15, 12489 Berlin, Germany}

\author{Peter J. \surname{Menck}}
\affiliation{Potsdam Institute for Climate Impact Research, P.O. Box 60 12 03, 14412 Potsdam, Germany}

\author{Jobst \surname{Heitzig}}
\affiliation{Potsdam Institute for Climate Impact Research, P.O. Box 60 12 03, 14412 Potsdam, Germany}

\author{J\"{u}rgen \surname{Kurths}}
\affiliation{Potsdam Institute for Climate Impact Research, P.O. Box 60 12 03, 14412 Potsdam, Germany}
\affiliation{Department of Physics, Humboldt University of Berlin, Newtonstr. 15, 12489 Berlin, Germany}
\affiliation{Institute for Complex Systems and Mathematical Biology, University of Aberdeen, Aberdeen AB24 3UE, United Kingdom}
\affiliation{Department of Control Theory, Nizhny Novgorod State University, 606950 Nizhny Novgorod, Russia}

\date{\today}

\begin{abstract} 
Stability assessment methods for dynamical systems have recently been complemented by
basin stability and derived measures, i.e. probabilistic statements whether systems remain in a basin of 
attraction given a distribution of perturbations. This requires numerical estimation via 
Monte-Carlo sampling and integration of differential equations. Here, we analyze 
the applicability of basin stability to systems with basin geometries challenging for 
this numerical method, having fractal basin boundaries and riddled or intermingled 
basins of attraction. We find that 
numerical basin stability estimation is still meaningful for fractal boundaries 
but reaches its limits for riddled basins with holes.
\end{abstract}



\maketitle

Going back to the path-breaking ideas of Aleksandr M. Lyapunov, 
dynamical systems are said to be stable if small variations of the initial conditions lead to 
small reactions of a system, i.e. small perturbations cannot substantially alter the system's state.
This is commonly a statement about the asymptotic behaviour, allowing for large transient 
deviations if only the system eventually returns to the initial state. 
\emph{Multistable} systems with several attractors add another subtlety to the problem: perturbations may lead to
switching from one attractor to another, substantially altering asymptotic behaviour \citep{Pisarchik2014}.
While infinitesimal perturbations on 
an attractor have local effects well-studied in the theory 
of asymptotic stability, finite (including large) perturbations can be critical by causing 
non-local effects like the transition to another attractor.

A direct stability method are Lyapunov functions 
\citep{Lyapunov1907,Hahn1958,Malisoff2009},
which decrease along trajectories and have local 
minima on attractors.
Finding Lyapunov functions is, however, difficult in high-dimensional 
multi-stable systems, although there are recent approaches to determine them from 
radial basis functions 
(e.g. \citep{Giesl2016}). 

Here, we pursue an alternative approach to consider non-local 
perturbations termed \emph{basin stability} $\bs$.
The central idea \citep{Menck2013,Menck2014} is to use a kind of volume of the basin of attraction 
to quantify the stability of attractors in multi-stable systems subject to a given distribution of
perturbations. An advantage of this measure is that it can be efficiently estimated even in 
high-dimensional systems 
and as an intuitive interpretation as a probability to return to an attractor, 
but it relies on the correct identification of the asymptotic behaviour for a Monte Carlo sample of initial conditions.
Basin stability and derived concepts have been successfully applied recently \citep{Rodrigues2016}, e.g. for power grids \citep{Kim2015,Kim2016,Menck2014,Schmietendorf2014}, chimera states \citep{Martens2016}, explosive synchronization \citep{Zou2014} delayed dynamics \citep{Leng2016} and resilience measures \citep{Mitra2015}.

In numerical simulations, it can be difficult to correctly identify the asymptotic behaviour and determine the
attractors. The basin of attraction can practically be defined as the set of all states that enter and stay in some trapping 
region \citep{Nusse1996}. Problems may arise if transients are long and chaotic or trajectories 
stay close to basin boundaries for long, 
so that numerical errors can move the simulated trajectory across a boundary into a wrong basin
and make the simulation converge to a wrong attractor.
Principally, three aspects contribute to the overall estimation error: the standard error due to sampling initial conditions,
approximation errors in function evaluations or integration of differential equations,
and rounding errors due to limited precision. 
While sampling and approximation errors are controlled by increasing sample size
and order of approximating polynomials and by decreasing step size, 
rounding errors are typically hard to reduce, 
which is not a problem if they are much smaller.

Our study thus focuses on the critical case of systems 
where rounding errors cannot be neglected and may even dominate the overall error
due to an intricate state space geometry highly sensitive to
numerical imprecisions. We put basin stability estimation here to the test by applying it to systems with 
fractal basin boundaries and riddled or intermingled
basins of attraction.

Consider a system of ordinary differential equations
\begin{equation}
    \dot{\vx} = \vF(\vx,t)
\end{equation}
that has more than one attractor in its state space $\mS$.
Here, we define an \emph{attractor} as a minimal compact invariant set $\mA \subseteq \mS$ whose basin of attraction 
has positive Lebesgue measure \citep{Milnor1985}. The {basin of attraction} of $\mA$ is the set $\mB(\mA) \subseteq \mS$ 
of all states from which the system converges to $\mA$.

Assume the system moves on an attractor $\mA$, 
yet at $t=0$ a random and not necessarily small perturbation 
pushes the system to a state $x(0)$ outside $\mA$.
Assume that $x(0)$ is drawn from a probability distribution with measure $\mu$ on $\mS$
that encodes our knowledge about what relevant perturbations are how likely to occur.
E.g., $\mu$ may be a uniform distribution on some bounded region
$\mR\supset\mA$.

{\em Will the system converge back to $\mA$ after the perturbation?}
To address this, we study the probability mass of $\mB$,
\begin{equation}
\bs\br{A} := \mu\br{\mB\br{A}} = \int\limits_\mR \ind{\mB\br{A}} ~d\mu \;\in [0;1], 
\label{eq2-bsdef}
\end{equation}
the probability that the system will return to $\mA$.
The indicator function $\ind{\mB\br{A}}\br{x}$ yields $1$ if $x\in \mB\br{A}$ 
and $0$ otherwise. 
We use $\bs\br{A}$ to quantify 
just \emph{how} stable the attractor $\mA$ is against non-infinitesimal perturbations,
and call it the \emph{basin stability} of $\mA$ \citep{Menck2013,Menck2014}.


The estimation of volume integrals such as Eq.\,\ref{eq2-bsdef} in high dimensions is a well-known problem, 
and we assume this is done by simple Monte-Carlo sampling \citep{Evans2000,Neumann1951}. 
If for each initial state $x(0)$, one can numerically integrate the system $x(t)$ 
with sufficient precision to decide to which attractor it converges or whether it diverges, the $\bs$ 
estimation procedure is thus:
\begin{enumerate}
    \item Draw a sample of $N>0$ independent initial states from the distribution $\mu$.
    \item For each, numerically integrate the system until it is clear whether and where it converges.
    \item Count the number $M$ of times the system has converged to $\mA$.
    \item Use the estimate $\hbs = \frac{M}{N}$.
\end{enumerate}
Since this is an $N$-times repeated Bernoulli experiment with success probability $\bs$,
the absolute standard error of the estimate $\hbs$ due to sampling is $\sqrt{\bs\br{A} (1-\bs\br{A})/N}$, 
independently of the system's dimension. Thus, the procedure can be applied to high-dimensional 
systems without necessarily increasing the sample size $N$, although it may take longer to assess convergence. 
This is of course since we are not interested in the basin of attraction's \emph{geometry} but only in 
its {\em volume} w.r.t.\ the measure $\mu$.

Note that when the relative std.\ err.\ of $\hbs$ is more relevant than the absolute std.\ err., 
smaller values of $\bs\br{A}$ require larger sample sizes, of the order $N\sim 1/\bs\br{A}$, 
since for small $\bs\br{A}$, the rel.\ std.\ err.\ is $\sim 1 / \sqrt{N \bs\br{A}}$.
However, even if $\bs\br{A}$ is not small, the geometries of the multiple basins of attraction may still make 
the estimation of $\bs$ difficult for another reason: For some initial conditions $x(0)$, it may be quite difficult to 
decide where $x(t)$ converges to, since the trajectory may start or come quite close to the boundary between the 
different basins, so that approximation and rounding errors (rather than sampling errors) in the integration may become relevant 
and may make the simulated trajectory hop across a basin border, leading to a wrong assessment of where $x(t)$ 
actually converges to.


Particularly, this is probable if the basins have \emph{fractal} boundaries, where 
the nature of the basin boundaries influences the predictability of a system's behaviour in the 
long run \citep{Grebogi1983a,McDonald1985,Nusse1996}. 
Imagine we randomly draw initial states from a box 
through which the boundary between the basins of two attractors runs. 
Suppose each initial state is specified up to a certain numerical error $\eps$. 
Then for an initial state that is closer to the boundary than $\eps$, 
it is uncertain to which of the two attractors the system will converge. 
Denote by $f(\eps)$ the fraction of initial states for which the outcome is uncertain,
i.e. the \emph{uncertainty fraction} \citep{Grebogi1983,Grebogi1987}. 
If the boundary is a smooth curve, then $f(\eps)$ is just proportional to $\eps$. 
However, if the boundary is fractal, then $f(\eps) \propto \eps^\alpha$.
If $\alpha<1$, the system exhibits \emph{final state sensitivity}, i.e., to decrease the 
uncertainty one needs a substantial improvement in the knowledge of initial conditions. 
In a way, this power law scaling leads to an \emph{obstruction of predictability} 
\citep{Grebogi1983} very similar to the sensitive dependence on initial conditions 
in chaotic systems. It has been found to occur in Rayleigh-B\'enard convection, 
e.g. numerically \citep{Lorenz1963} and experimentally \citep{Berge1983}. 

Predicting the long-term behaviour -- the essence of estimating $\bs\br{A}$ -- 
of systems with fractal basin boundaries may be hard \citep{McDonald1985} although 
generally, for most initial conditions, the final state sensitivity  
is much smaller than the unpredictability of the actual trajectory.

\begin{figure}[tpb]
\centering
\includegraphics[width=\columnwidth]{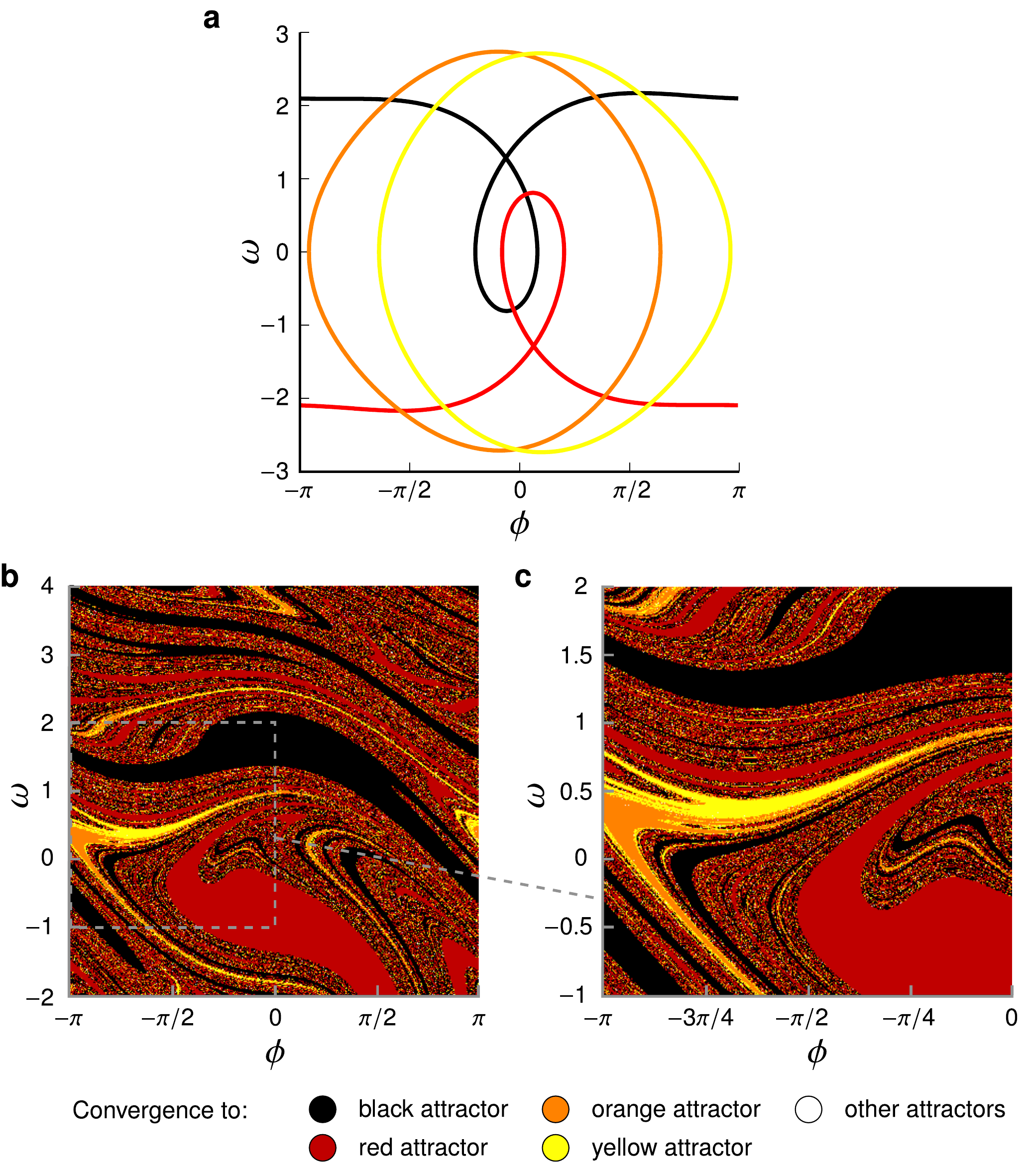}
\caption[Damped pendulum with fractal basin boundaries.]{
	(Color online)
	\textbf{Damped pendulum with fractal basin boundaries.}
	Damped pendulum with fractal basin boundaries. 
	\textbf{(a)} Attractors of the damped pendulum with time-dependent forcing from Eqn.\ (\ref{eq2-swing}). 
	\textbf{(b)} State space of the pendulum at $t=0$. 
	Black/red/orange/yellow colouring indicates convergence to the black/red/orange/yellow attractor. 
	Convergence to other attractors is indicated by white colouring. 
	\textbf{(c)} Detail of dashed square from (b).
}
\label{fig:f1}
\end{figure}

\begin{figure}[tpb]
\centering
\includegraphics[width=\columnwidth]{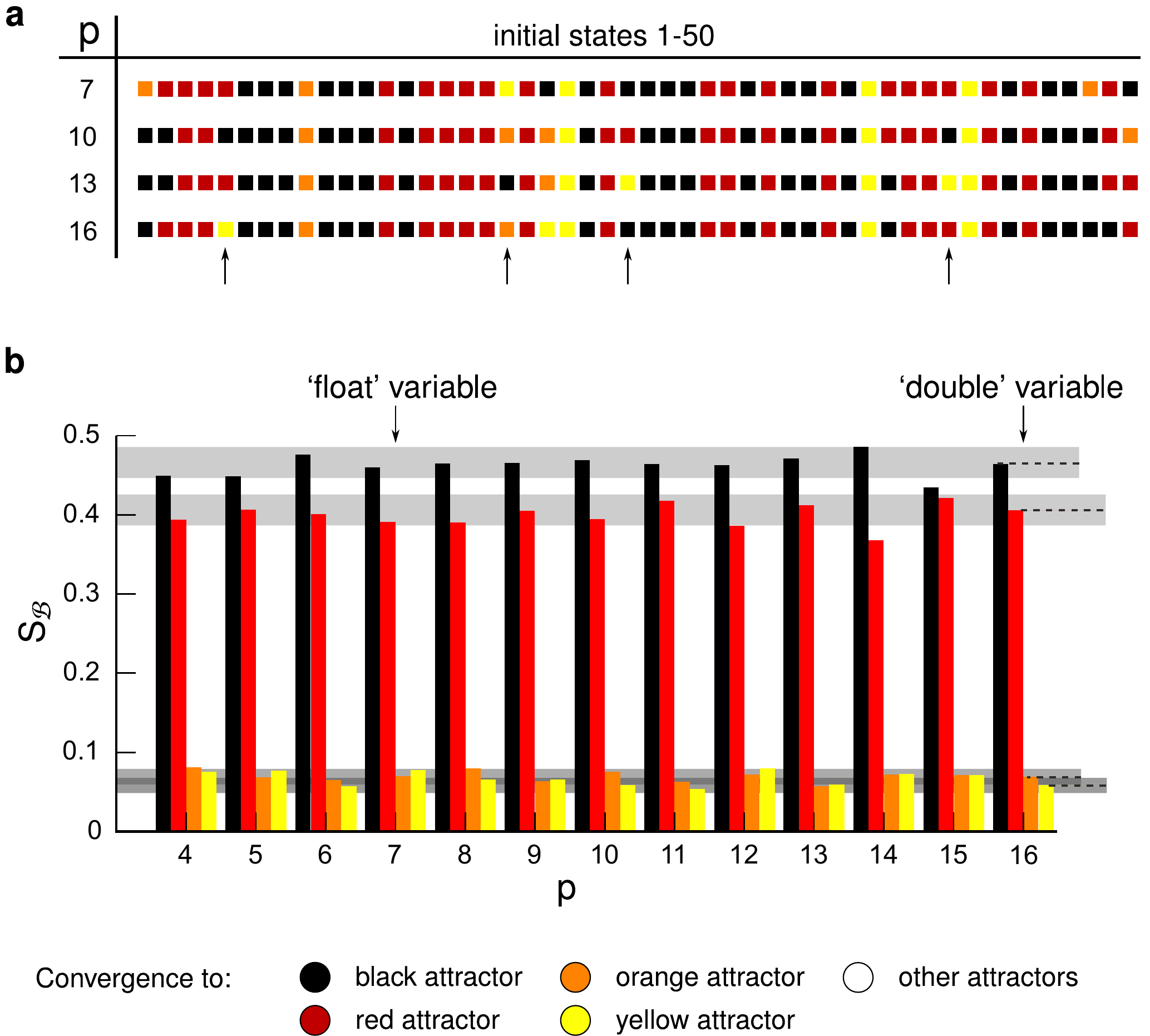}
\caption[Basin stability in the pendulum with fractal basin boundaries.]{
	(Color online)
	\textbf{Basin stability in the pendulum with fractal basin boundaries.}
    \textbf{(a)} Numerical integrations for a fixed set of fifty initial states
    at different values of the numerical precision $p$. 
    The squares in each column correspond to the same initial state, 
    and their respective colours indicate which state the system converges to from there at given precision $p$. 
    Black/red/orange/yellow colouring indicates convergence to the black/red/orange/yellow attractor. 
    \textbf{(b)} Estimated basin stability $\hbs$ of the four attractors at different levels of $p$ using $N=1000$. 
    The basin stability of the black/red/orange/yellow attractor 
    is shown by the height of the black/red/orange/yellow bar. 
    The grey shadows indicate the standard error of $\hbs(p=16)$ .
}
\label{fig:f2}
\end{figure}

So let us first investigate how fractal basin boundaries impact the accuracy of $\hbs$
by studying the \emph{Wada pendulum} \citep{Battelino1988,Grebogi1988}.
Consider a damped, driven pendulum that is subject to a time-dependent forcing:

\begin{equation}
\dot{\phi} = \omega,
\hspace{0.3cm}
\dot{\omega} = X \cos t - \alpha \omega - K \sin \phi \;.
    \label{eq2-swing}
\end{equation}

For $\alpha=0.1$, $K=1$ and $X=7/4$, this system  has several attractors \citep{Kennedy1991}.
The four dominant of them, all limit cycles with period $2\pi$, are shown in Fig.~\ref{fig:f1}a: 
The black and red attractors correspond to rotations of the pendulum, 
and the orange and yellow attractors are librations. 
Their respective basins of attraction at $t=0$ are shown in Fig.~\ref{fig:f1}b. 
Certain regions in this figure appear sprinkled with dots belonging to the different basins, 
i.e. the boundary between the basins is not easily discernible and remains so when zooming 
in (Fig.~\ref{fig:f1}c). It is a fractal, resulting from the so-called Wada property of the basins.

Three (or more) subsets of a space are said to have the \emph{Wada property} if any point on the boundary of one 
subset is also on the boundary of the two others \citep{Kennedy1991, Nusse1996}.
For the pendulum, the black basin, the red basin and the union of the orange and yellow basins 
have the Wada property \citep{Kennedy1991, Nusse1996}. This means that starting within the rounding error $\eps$ 
of the boundary, a trajectory could in principle converge to \emph{any of the four} attractors.

To verify this empirically, we write $\eps = 10^{-p}$ with $p$ denoting \emph{precision}, 
and discard all information after the $p$-th significant decimal digit in the floating point variables 
used in all individual operations of the numerical integration. We use 64 bit double precision to allow for 
a maximum of $p=16$, while using untruncated 32 bit single precision would correspond to $p\approx 7$. 
For different values of $p$, we integrate a fixed set of 50 initial states $x(0)$, drawn uniformly at random from the rectangle
$\mR = [-\pi, \pi] \times [-2,4]$. 

Fig.~\ref{fig:f2}a, reveals that some initial states, particularly those indicated by arrows, 
indeed lead to different outcomes for different values of $p$.
To investigate how $\hbs$ depends on $p$, we let $\mu$ be the uniform distribution on $\mR$ 
yielding a sample of $N=1,000$ random initial states which are integrated with different precisions $p$, 
leading to estimates $\hbs(p)$.
As depicted in Fig.~\ref{fig:f2}b, there seems to be no systematic influence of $p$ on $\hbs(p)$. 
Indeed, most of the individual values of $\hbs(p)$ are within one standard error of the most precise value $\hbs(16)$. 
This suggests that, in contrast to long-term prediction for {\em individual} initial states (cf.\ Fig.~\ref{fig:f2}a), 
$\hbs$ is robust under variation of $p$.

\begin{figure}[tpb]
\centering
\includegraphics[width=\columnwidth]{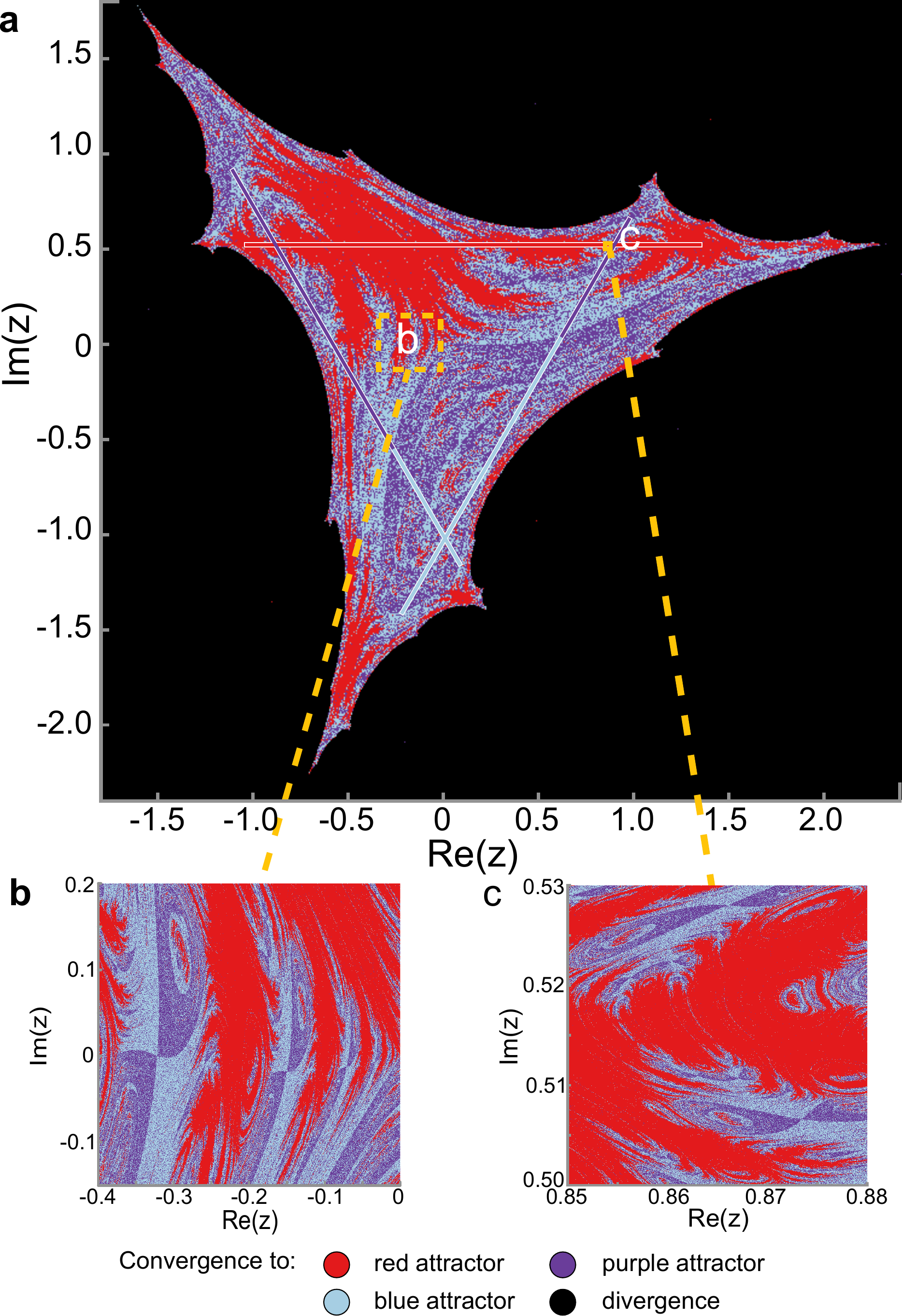}
\caption[Intermingled basins of the quadratic map.]{
	(Color online)
	\textbf{Intermingled basins of the quadratic map.}
	\textbf{(a)} Phase space portrait of the three attractors (red/blue/purple line segments) of the map (Eqn.~\ref{eqn:quad}) 
	with their intermingled basins of attraction coloured alike. The black area corresponds to initial conditions for which the 
	dynamics diverge. Below are zoom-ins of two regions, \textbf{(b)}and \textbf{(c)}. The locations of the attractors
	(line segments, see \citep{Alexander1992}) are 	highlighted by red/blue/purple bars (not in scale).}
\label{fig:f3}
\end{figure}

\begin{figure}[tpb]
\centering
\includegraphics[width=\columnwidth]{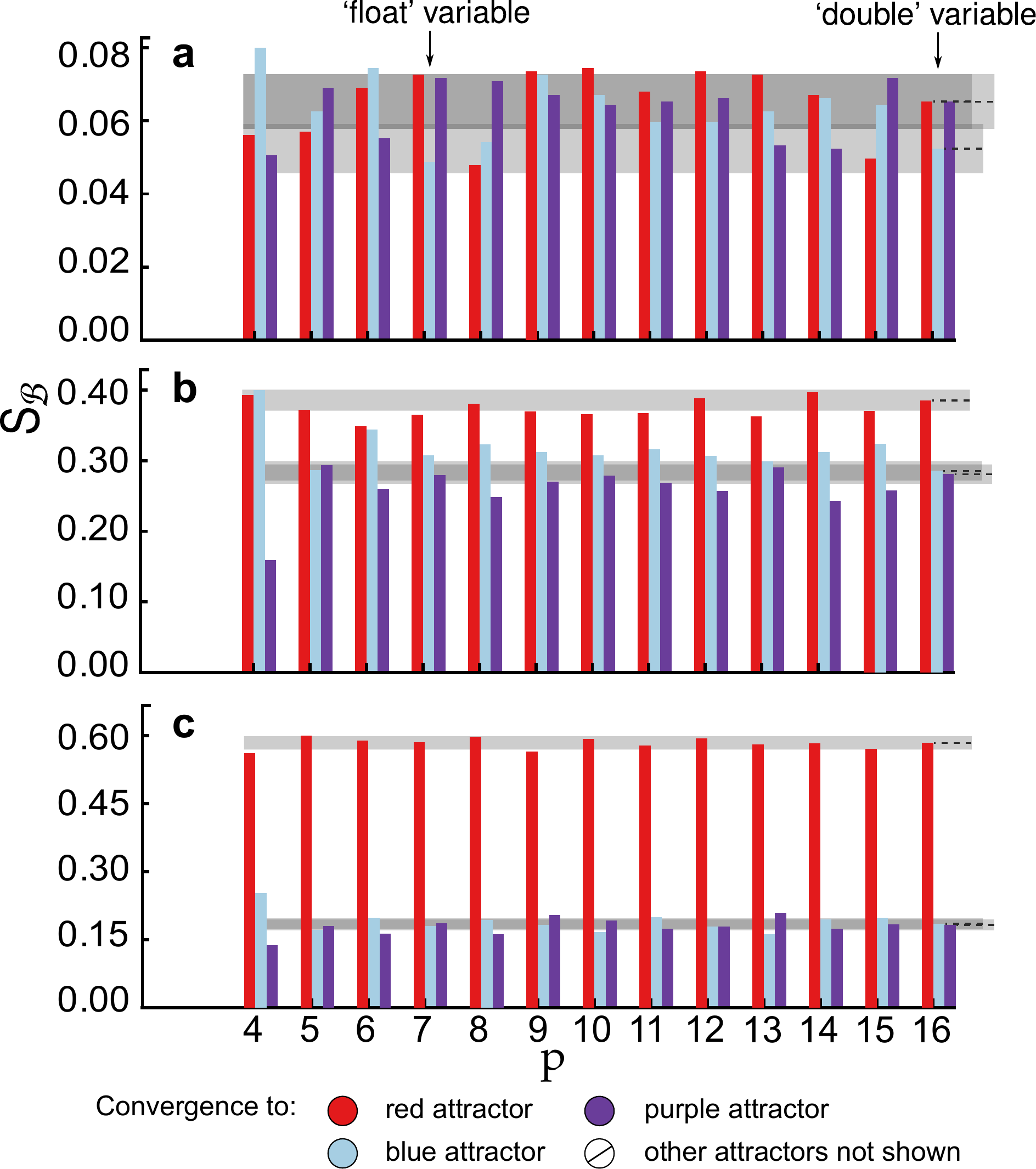}
\caption[Basin stability estimation for the quadratic map.]{
	(Color online)
	\textbf{Basin stability estimation for the quadratic map.}
	\textbf{(a)} $\hbs$ of the the red/blue/purple attractors at different levels of $p$, using $\mR = [-1.8, 2.4] \times [-2.4,1.8]$.
	\textbf{(b)} $\hbs$ with  $\mR$ corresponding to Fig.~\ref{fig:f3} inset (1), 	
	\textbf{(c)} $\hbs$ with $\mR$ corresponding to Fig.~\ref{fig:f3} inset (2).
    The basin stability is shown by the height of the red/blue/purple bar, the grey shadows indicate the standard error of
    $\hbs(16)$.
}
\label{fig:f4}
\end{figure}

Another extreme case are attractors whose basins are not open as for most systems \citep{Milnor1985} but 
rather have an empty interior. The complement of such a \emph{riddled basin}
 intersects every disk in a set of positive measure \citep{Alexander1992,Ott1994,Lai1996,Lai1996a}. This means that \emph{all} points 
 in its basin of attraction have pieces of \emph{another} attractor basin arbitrarily closely nearby \citep{Ott1994}.

Physical systems exhibiting riddled basins are the damped, periodically-driven particle moving in a special potential \citep{Sommerer1993} or coupled time-delayed systems \citep{Ashwin2005,Chaudhuri2014, Jiang2000}. 
There are also experimental observations for laser-cooled ions in a Paul trap \citep{Shen2008} indicating a 
riddled phase space structure.

In the following, we investigate the impact of riddled basins of attraction on $\hbs$
using a conceptual example \citep{Alexander1992,Lopes1992}, i.e. the following quadratic map 
on the complex plane:

\begin{align}
F_\lambda (z) = z^2 - (1 + \lambda i) \bar z , \hspace{0.3cm} \lambda = 1.02871376822 \,.  \label{eqn:quad}
\end{align}

This map has three different attractors on the complex plane which are shown in Fig.~\ref{fig:f3}; for simplicity
they are referred to as the red/blue/purple attractors with their respective basin of attraction  in the following. 
Interestingly, the three basins of attraction are not just riddled, they are \emph{intermingled}.
A basin of attraction is called intermingled if \emph{any} open set which intersects one basin in a set of positive 
measure also intersects \emph{each} of the other basins in a set of positive measure \citep{Kan1994,Lai1995}. 

The fact that there is a positive probability to end up in a different attractor around each initial condition
inside a riddled/intermingled basin of attraction renders these systems effectively non-deterministic \citep{Sommerer1993}.
As in the case of Wada boundaries, slight variations of initial conditions or numerical imprecisions will affect any 
forecast of the system's long-term behaviour.

Again, we investigate the effect of limited numerical precision on the significance of $\hbs$.
In Fig.~\ref{fig:f4}a we depict the result of estimating $\bs$ for varying $p$ using $\mR = [-1.8, 2.4] \times [-2.4,1.8]$, 
i.e. the region pictured in Fig.~\ref{fig:f3}a. We observe a large variation of $\hbs$ of up to $50\%$  
compared to the most precise estimation $\hbs(16)$ and no systematic dependence on $p$. 

In Fig.~\ref{fig:f3}c we zoomed into the neighbourhood of the red attractor, where the 
share of the corresponding red basin is increasing in proximity of the attractor.  In particular, the measure of 
this basin of attraction, restricted to an $\epsilon$-neighbourhood of the attractor,  approaches unit probability 
for $\epsilon\to 0$ \citep{Alexander1992}. This apparent behaviour provides an explanation for Fig.~\ref{fig:f4}c where we 
determined $\hbs(p)$ for Fig.~\ref{fig:f3}c. In contrast to our previous 
observation, the fluctuations of  $\hbs(p)$ almost stay within one standard error and the estimation appears to be 
more robust. For reference, Fig.~\ref{fig:f4}b depicts $\hbs(p)$ for Fig.~\ref{fig:f3}b
not containing any (part of) an attractor. On the one hand, the variation of $\hbs(p)$ exceeds one standard error, up to 
about $20\%$ compared to $\hbs(16)$, such that our estimation is more sensitive to numerical imprecisions
than in Fig.~\ref{fig:f4}b; on the other hand the variations are smaller than in our first experiment. 

In conclusion, we applied the Monte-Carlo estimation procedure of basin stability in two cases, i.e. 
basins with fractal boundaries and riddled/intermingled basins of attraction. 
In the former case, we find that while the asymptotic properties of individual trajectories still cannot be determined robustly, 
the converse is true for the basin stability estimation. It remains an open question for future research, how exactly (in a quantitative
sense) the numerical estimation uncertainty might be derived from the actual basin geometry. 
In the latter case, however, we find that the results can vary drastically with the chosen precision. 
The effect of rounding errors is comparable or even larger than the standard error of the sampling. Only if the sample 
region $\mR$ is chosen in some sense "close enough" to the actual attractor of interest, the foliated structure of the 
surrounding basins allows for a meaningful numerical estimation.

\emph{What are practical implications for the application of basin stability? }
In general,  it is sufficient if the rounding error of an estimation is smaller than its sampling error to get a significant result.
However, any numerical procedure is subject to a finite numerical precision and we have to assume that in practice it will not be 
high enough to reach this goal in dynamical systems with intricate basin geometries. If there is no prior knowledge available,
a good starting point is to actually visualize the interesting part of the phase space to get a first idea of the appearance of, e.g., 
fractal sets. If any are detected, it is necessary to use the highest available numerical precision $p_h$ to get $\hbs(p_h)$, 
potentially avoiding artifacts respectively insignificant estimations. We 
suggest to repeat the $\bs$ estimation at a lower numerical precision $p_l$ and take the difference $\hat{e}_p = 
\vert\hbs(p_h)-\hbs(p_l)\vert$ as a straight-forward (rough) estimator of the variability of $\hbs(p)$ with $p$ and, by way of
extrapolation, as a rough estimate of the remaining standard error of $\hbs(p_h)$ as an estimate of $\bs$ due to finite numerical 
precision. To assess the influence of rounding errors on $\hbs$ then compare $\hat{e}_p$ with the 
standard error of $\hbs(p_h)$ as an estimate of $\bs(p_h)$ due to sampling, which can be estimated as 
$\hat{s}_p = \sqrt{\hbs(p_h)(1-\hbs(p_h))/N}$. If $\hat{e}_p < \hat{s}_p$, rounding has no significant effect on the estimation 
quality. For instance, this could be implemented by comparing the results at double and single precision computations. 

The authors gratefully acknowledge the support of BMBF, CoNDyNet, FK. 03SF0472A.


%

\end{document}